*Original Paper*

# Comment on Causal Networks and Freedom of Choice in Bell's Theorem

**Marian Kupczynski**

Département d'informatique et d'ingénierie, Université du Québec en Outaouais, Gatineau, QC, Canada
*E-Mail: marian.kupczynski@uqo.ca*

**Abstract:** Bell inequalities may only be derived, if hidden variables do not depend on the experimental settings. The stochastic independence of hidden and setting variables is called: freedom of choice, free will, measurement independence (MI) or no conspiracy. By imbedding the Bell causal structure in a larger causal network the authors correctly prove, that one can explain and quantify possible violation of MI without evoking super-determinism. They assume the independence of the variables that causally determine the settings and investigate how they might become correlated with hidden variables (e.g., when the cosmic photons enter the laboratory). Using their extended causal networks they derive a contextual probabilistic model on which their further correct results are based. The authors seem to ignore that contextual probabilistic model may be derived directly using only probabilistic concepts and incorporating correctly setting dependent variables describing measuring instruments. In these contextual probabilistic models experimenters' freedom of choice is not compromised and the results of Bell Tests including an apparent violation of Einsteinian non-signaling may be explained in a locally causal way. Talking about freedom of choice is misleading and is rooted in incorrect understanding of Bayes Theorem. We explain why MI should be called noncontextuality and why its violation in Bell Tests confirms only the contextual character of quantum observables. Therefore, contextuality and not experimenters' freedom of choice are important resources in quantum information.

In Bell scenario we have 4 random experiments performed using 4 incompatible pairs of settings [1, 2]. A source S is sending two correlated photonic signals to Alice and Bob in distant laboratories. Before 'photons' arrive to corresponding measuring instruments Alice and Bob choose independently their experimental settings (x, y). In quantum mechanics (QM) and in Bell model, x=1 or 2 and y=1 or 2 are labels indicating which pair of experimental setting are chosen. In Bell tests (x, y) are values of two binary random variables (X, Y) describing outcomes of supplementary random experiments (e.g. flipping two fair coins, generation of pseudo-random numbers) or coded information from some random phenomena (e.g. arrivals of cosmic photons [3-4], unpredictable bits created by smartphones and other internet-connected devices [5]. After passing by distant polarization beam splitters the 'photons' produce two time-series of clicks on distant detectors coded by corresponding random variables: $A_x = \pm 1$ and $B_y = \pm 1$.

In Bell 64 [1,2] local realistic hidden variable model (LRHVM), a source is producing a mixed statistical ensemble $\Lambda$ of particle pairs carrying strictly correlated properties $\lambda$ and outcomes $\pm 1$ are predetermined by these properties for all the settings. The randomness of the outcomes is created at the source and not during the measurement:

$$E(A_x B_y) = \sum_{\lambda \in \Lambda} A_x(\lambda) B_y(\lambda) p(\lambda) \qquad (1)$$

where $A_x(\lambda) = \pm 1$ and $B_y(\lambda) = \pm 1$. According to the model (1), there exists a counterfactual joint probability distribution (JP) of random variables describing the outcomes obtained in all different experimental setting. In particular:

$$E(A_x A_{x'} B_y B_{y'}) = \sum_{\lambda \in \Lambda} A_x(\lambda) A_{x'}(\lambda) B_y(\lambda) B_{y'}(\lambda) p(\lambda) \qquad (2)$$

Using (1) and (2) one derives easily Clauser-Horne-Shimony–Holt (CHSH) inequalities [6]:

$$| E(A_x B_y) + E(A_x B_{y'}) + E(A_{x'} B_y) - E(A_{x'} B_{y'}) | \leq 2 . \qquad (3)$$

As Fine demonstrated [7, 8], CHSH are necessary and sufficient conditions for the existence of a counterfactual JP of $A_x$, $A_{x'}$, $B_y$ and $B_{y'}$.

In LRHVM [1], $\lambda$ describe only entangled 'particle pairs', thus it is reasonable to assume that setting variables and hidden variables in (1) are stochastically independent:

$$p(x, y, \lambda) = p(x, y) p(\lambda), \ p(x\, y| \lambda) = p(x, y), \ p(\lambda | x, y) = p(\lambda.) \qquad (4)$$

The stochastic independence (4) is called: *no-conspiracy*, *freedom of choice*, *free will* or *measurement independence* (MI) [9]. Since the inequalities (3) are violated for some settings by quantum predictions and by experimental data, it has been debated which assumptions in the model (1) might be relaxed. One might relax locality (e.g. $A_x = A_x(\lambda, y)$ and $B_y(\lambda, x)$) or relax MI [10, 11, 12] and assume that: $p(\lambda | x, y) \neq p(\lambda.$

However, the relaxation of locality would allow for spooky influences between distant measuring instruments and $p(x\, y | \lambda) \neq p(x, y)$ is often believed to imply the *super-determinism* [13]. Such beliefs are unjustified. As we explained in [14,15], conditional probabilities in Bayes Theorem do not admit, in general, any causal interpretation and the violation of MI does not constraint *experimenters' freedom of choice*, which is in fact the prerequisite of science.

We also explained, that MI should be called *noncontextuality*, because it allows to describe random experiments performed in incompatible experimental settings using a counterfactual joint probability distribution (JP) on a unique probability space $\Lambda$. The violation of Bell inequalities proves only that pairwise expectations in the inequality (3) are not marginal expectations derived from a non-existing JP [16, 17]. CHSH are simply *noncontextuality inequalities* for 4-cyclic scenarios [18].

By embedding the Bell causal structure in a larger causal network, Rafael Chavez et al. [12] prove, that it is not necessary to evoke *super-determinism* in order to justify the violation of MI. They assume the independence of the variables that causally determine the settings and investigate how they might become correlated with hidden variables (e.g., when the cosmic photons enter the laboratory [3, 4]). Using their extended causal networks they derive a contextual probabilistic hidden variable model on which their further correct results are based:

$$p(a,b|x,y) = \sum_\lambda p(a|x,\lambda)p(b|y,\lambda)p(\lambda|x,y) \qquad (5)$$

However, they do not realize that calling the statistical independence (4) *freedom of choice* is misleading. They also seem to ignore, that that the simplest way to justify the contextual probabilistic model (5) is the inclusion of local setting dependent hidden variables describing the measuring instruments at the moment of the their interaction with the correlated photonic signals arriving from the source.

Such models were proposed several years ago and have been recently updated and reviewed extensively in [15,17]. They provide a locally causal explanation not only of imperfect correlations between distant clicks, but also the explanation of an apparent violation of Einsteinian non-signalling discovered in the data of several Bell Tests [19-22].

Let us consider a spin polarisation correlation experiment (SPCE) using a protocol similar to that of Weihs et al. [23]. Alice and Bob have each two detectors and register time series of clicks. From raw data one has to extract final samples in order to estimate correlations and perform a Bell test. A detailed discussion, how it is done may be found in [19, 20, 23, 24]. As we explained in [14]:

1. Raw time-tagged data are two samples: $S_A(x, y) = \{(a_k, t_k) | k=1…n_x\}$ and $S_B(x, y) = \{(b_m, t'_m) |j=1…n_y\}$ with $a_k = \pm 1$ and $b_m = \pm 1$.

2. Using fixed synchronized time-windows of width W and keeping only time-windows in which there is no click at all or a click on one of Alice's or/and Bob's detectors two new samples are created: $S_A(x, y, W) = \{a_s | s=1,…N_x\}$, $S_B(x, y, W) = \{b_t | t = 1…N_y\}$ with $a_s = 0, \pm 1$ and $b_t = 0, \pm 1$

3. Now by keeping only synchronized time-windows, in which both Alice and Bob observed a click on one of their detectors, new sample of paired outcomes is obtained: $S'_{AB}(x, y, W) = \{(a_r, b_r) | r=1,…N_{xy}\}$ with $a_r = \pm 1$ and $b_r = \pm 1$.

Samples, obtained at the step 2, may be described by the following non-signalling contextual hidden variable model

$$E(A_x B_y) = \sum_{\lambda \in \Lambda_{xy}} A_x(\lambda_1, \lambda_x) B_y(\lambda_2, \lambda_y) p_x(\lambda_x) p_y(\lambda_y) p(\lambda_1, \lambda_2) \qquad (6)$$

where $A_x(\lambda_1, \lambda_x) = 0, \pm 1$, $B_y(\lambda_2, \lambda_y) = 0, \pm 1$. We have 4 random variables $(L_1, L_2, L_x, L_y)$ taking values $\lambda = (\lambda_1, \lambda_2, \lambda_x, \lambda_y)$, where $(\lambda_1, \lambda_2)$ describe photonic signals and $(\lambda_x, \lambda_y)$ the measuring instruments, how they are 'perceived' by the signals at the moment of interaction. Four incompatible experiments are described by probability distributions defined on disjoint hidden variable spaces:

$$\Lambda_{xy} = \Lambda_{12} \times \Lambda_x \times \Lambda_y; \Lambda_{x'y} = \Lambda_{12} \times \Lambda_{x'} \times \Lambda_y; \Lambda_{xy'} = \Lambda_{12} \times \Lambda_x \times \Lambda_{y'}; \Lambda_{x'y'} = \Lambda_{12} \times \Lambda_{x'} \times \Lambda_{y'} \quad (7)$$

where $\Lambda_x \cap \Lambda_{x'} = \Lambda_y \cap \Lambda_{y'} = \emptyset$. Probability distributions of hidden variables depend on the settings: $p(\lambda | x, y) = p_x(\lambda_x) p_y(\lambda_y)$ and:

$$p(a, b | x, y) = \sum_{\lambda \in \Lambda_{ab}} p_x(\lambda_x) p_y(\lambda_y) p(\lambda_1, \lambda_2) \qquad (8)$$

where $\Lambda_{ab} = \{\lambda \in \Lambda_{xy} | A_x(\lambda_1, \lambda_x) = a$ and $B_x(\lambda_2, \lambda_y) = b\}$. Using the model (6-8) one may not derive CHSH.

In order to perform Bell Test (3) the samples obtained in the step 3 have to be used. From (6-8): we easily derive required expectations:

$$E(A_x B_y \mid A_x B_y \neq 0) = \sum_{\lambda \in \Lambda'_{xy}} A_x(\lambda_1, \lambda_x) B_y(\lambda_2, \lambda_y) p_{xy}(\lambda) \qquad (9)$$

$$E(A_x \mid A_x B_y \neq 0) = \sum_{\lambda \in \Lambda'_{xy}} A_x(\lambda_1, \lambda_x) p_{xy}(\lambda) \qquad (10)$$

$$E(B_y \mid A_x B_y \neq 0) = \sum_{\lambda \in \Lambda'_{xy}} B_y(\lambda_1, \lambda_y) p_{xy}(\lambda) \qquad (11)$$

where $p_{xy}(\lambda) = p_x(\lambda_x) p_y(\lambda_y) p(\lambda_1, \lambda_2)$, $\Lambda_{xy} = \Lambda_1 \times \Lambda_2 \times \Lambda_x \times \Lambda_y$ and

$$\Lambda'_{xy} = \{\lambda \in \Lambda_{xy} \mid A_x(\lambda_1, \lambda_x) \neq 0, B_y(\lambda_2, \lambda_y) \neq 0\}. \qquad (12)$$

The model (9-12) allows explaining the reported violation of Einsteinian non-signalling:

$$E(A_x \mid A_x B_y \neq 0) \neq E(A_x \mid A_x B_{y'} \neq 0); E(B_y \mid A_x B_y \neq 0) \neq E(B_y \mid A_{x'} B_y \neq 0) \qquad (13)$$

This apparent violation of non-signalling does not prove the nonlocality of Nature, it is the result of the setting dependent post-selection (PSL) of events, such that both Alice and Bob register clicks on one of their detectors in corresponding synchronized time-windows.

The model (6-13) contains enough free parameters to fit experimental data [17] and to reproduce any violation of the inequality (3). The violation of CHSH inequality is not surprising, because it may not be derived for the pairwise expectations defined by (9).

The inequalities (13) prove that random variables describing the data obtained in the step 3 are inconsistently connected, thus a rigorous description of Bell Test should be done using the *Contextuality –by- Default* (CbD) approach of Dzhafarov and Kujala [25-28]. In CbD, random variables measuring the same content in different contexts are *stochastically unrelated* and they are labelled by contexts in which they are measured.

Therefore in Bell Tests we do not have a 4-cyclic system X= { $A_x$, $A_{x'}$, $B_x$, $B_{x'}$ } of 4 random variables but a system X' = {$A_{xy}$, $A_{xy'}$, $B_{xy}$, $B_{x'y}$, $A_{x'y}$, $A_{x'y'}$, $B_{xy'}$, $B_{xy}$} of 8 binary inconsistently connected random variables : $E(A_{xy}) \neq E(A_{xy'}), \ldots E(B_{xy'}) \neq E(B_{x'y'})$ [15]. The only inequality which may be derived for these random variables without additional assumptions is:

$$S' = E(A_{xy} B_{xy}) + E(A_{xy'} B_{xy'}) + E(A_{x'y} B_{x'y}) - E(A_{x'y'} B_{x'y'}) \leq 4 \qquad (14)$$

where $E(A_{xy} B_{xy}) = E(A_x B_y \mid A_x B_y \neq 0)$ etc. By imposing maximal couplings between observables measuring the same content in different contexts one derives more restrictive inequalities, which may be compared with the experimental data [15].

There is an intimate relation between probabilistic models and experimental protocols [16, 29]. Let us analyze this relationship in the contextual model (6-8). For each setting (x, y), we have 4 random variables (X,Y,$A_x$,$B_y$) describing observed outcomes; (x, y, $a_x$, $b_y$) and 4 hidden random variables:

($L_1, L_2, L_x, L_y$) taking values ($\lambda_1, \lambda_2, \lambda_x, \lambda_y$), The experimental protocol may be subdivided into the following "random experiments":

1. Generate variables $(\lambda_1, \lambda_2) \in \Lambda_{12}$ describing and an entangled pair.
1. Generate labels (x, y).
2. Generate variables $\lambda_x \in \Lambda_x$ and $\lambda_y \in \Lambda_y$ describing the instruments.
3. Evaluate: $a_x = A_x(\lambda_1, \lambda_x)$ and $b_y = B_y(\lambda_2, \lambda_y)$.
4. Output: x, y, $a_x$, $b_y$.

The setting variables X, Y are causally independent from **all** other random variables. They also do not stochastically depend on $L_1$ and $L_2$. The variables $L_1$ and $L_2$ are stochastically dependent. The variables $A_x$, $B_y$ are causally independent, but are stochastically dependent. The setting variables (X, Y) and instrument variables ($L_x$, $L_y$) are also stochastically dependent:

$$p(\lambda) = p_x(\lambda_x) p_y(\lambda_y) p(x, y) p(\lambda_1, \lambda_2) = p(x, y, \lambda) \Rightarrow p(x, y | \lambda) = \frac{p(x, y, \lambda)}{p(\lambda)} = 1 \qquad (15)$$

It does not mean that the experimenters' freedom of choice is compromised. It means only: if a "hidden event" $\{\lambda_1, \lambda_2, \lambda_x, \lambda_y\}$ "happened", thus the settings (x, y) were chosen.

Therefore, the stochastic independence of setting and hidden variables (4) should be called *noncontextuality* and not *freedom of choice*. A much more detailed discussion of MI, Bayes Theorem and its implications may be found in [14].

Misleading terminology is a source of speculations about *super-determinism*, *retro-causation*, and *quantum nonlocality* not only on the social media but also in serious scientific journals. It is high time to understand that adding the instrument variables is the simplest explanation of statistical dependence of setting and hidden variables.

Some incorrectly believe that Bell rejected this solution in 1970, because after averaging over instrument variables in (6), CHSH inequalities may still be derived [30]. However he did not realized, that after such averaging the resulting probabilistic model describes different random experiments with experimental protocols which are impossible to implement [14, 17, 29].

Long time ago, several authors explained, that the violation of Bell inequalities proved only, that incompatible spin polarization correlation experiments do not allow for a *noncontextual* description in terms of a joint probability on a unique probability space e.g. [7, 8, 16. 31-37].
*We* cited here only few papers more references may be found in [17] and in the papers cited therein.

Bell inequalities were proven assuming *predetermination of experimental results*, called also *local realism* or *counterfactual definiteness* (CFD). The inequalities are violated, thus as Peres [38] resumed: *unperformed experiments have no results*.

The results do not exist because in quantum mechanics measuring instruments play an active role. Nieuwenhuizen [39.40] pointed out that LRHVM suffered from incurable *contextuality loophole*, because this model did not include correctly variables describing measuring instruments.

Some authors claim that inequalities are violated because of *quantum nonlocality*. Khrennikov [41] clearly explained that *quantum nonlocality* is a misleading notion, which should be abandoned.

Therefore the important resource in quantum information is not the violation of *freedom of choice*, as it was called in [11,12], but *contextuality*. It is consistent with the contextual character of quantum observables, Bohr's complementarity [44] and *KS-contextuality* [45].

*Experimenters' freedom of choice* is the prerequisite of science and the violation of inequalities in various Bell Tests does not allow doubting in its validity.

*International Journal of Quantum Foundations* **8** (2022)